\documentclass[]{spie}  
\usepackage[]{graphicx}
\usepackage[]{subfigure}
\title{ A photoelectric polarimeter for XEUS: a new window in x-ray sky }

\author{ Ronaldo Bellazzini\supit{a},
Luca Baldini\supit{a}, Francesco Bitti\supit{a}, Alessandro
Brez\supit{a}, Francesco Cavalca\supit{a}, Luca
Latronico\supit{a},  Marco Maria Massai\supit{a},  Nicola
Omodei\supit{a}, Michele Pinchera\supit{a}, Carmelo
Sgr{\'o}\supit{a}, Gloria Spandre\supit{a}, Enrico Costa\supit{b},
Paolo Soffitta\supit{b}, Giuseppe Di Persio\supit{b}, Marco
Feroci\supit{b},    Fabio  Muleri\supit{b},  Luigi
Pacciani\supit{b}, Alda  Rubini\supit{b}, Ennio Morelli\supit{c},
Giorgio Matt\supit{d}, Giuseppe Cesare Perola\supit{d}
\skiplinehalf \supit{a} Istituto Nazionale di Fisica Nucleare,
Largo B. Pontecorvo 3, I-56127 Pisa,  Italy
\\
\supit{b} Istituto di Astrofisica Spaziale e Fisica Cosmica, Via
del Fosso del Cavaliere 100, I-00133 Roma, Italy;
\\
\supit{c} Istituto di Astrofisica Spaziale e Fisica Cosmica, Via
Gobetti 101, I-40129 Bologna, Italy
\\
\supit{d} Universita' degli Studi di Roma 3, Via della Vasca
Navale 84, I-00146 Roma, Italy }

\authorinfo{Further author information: (Send correspondence to Ronaldo Bellazzini)
\\R.Bellazzini: E-mail: ronaldo.bellazzini@pi.infn.it, Telephone: +39 050 2214367 }

\begin{document}
  \maketitle

\begin{abstract}
XEUS is a large area telescope aiming to rise X-ray Astronomy to
the level of Optical Astronomy in terms of collecting areas. It
will be based on two satellites, locked on a formation flight, one
with the optics, one with the focal plane. The present design of
the focal plane foresees, as an auxiliary instrument, the
inclusion of a Polarimeter based on a Micropattern Chamber. We
show how such a device is capable to solve open problems on many
classes of High Energy Astrophysics objects and to use X-ray
sources as a laboratory for a substantial progress on Fundamental
Physics.
\end{abstract}



\section{INTRODUCTION}
\label{sect:intro}  

If we look back to the history of X-ray Astronomy we can say that
after the first pioneering activity with rockets the sequence
UHURU, Einstein, Chandra, following a path tracked by R. Giacconi,
has been the backbone of this activity, that nowadays has arrived
to a turning point. No major mission is presently established and
ideas about how such a mission should be designed are not clear.
Chandra has shown the great capabilities of an optics at
sub-arcsecond level. Both Chandra and Newton have opened the
window of high resolution spectroscopy. In some sense these two
missions have brought X-ray Astronomy much closer to Optical
Astronomy, but when we convert the effective areas of these
telescopes into equivalent diameters we see that from the
collection capability, they are equivalent to medium level optical
amatorial telescopes, although equipped with an extremely
performing focal plane instrumentations.

These rose the need of a new generation telescope with a
collecting area two orders of magnitude higher, while still
preserving an angular resolution at arcsecond level, which is
imperative to study crowded deep fields avoiding source confusion.
The availability of International Space Station suggested that
such a telescope could be assembled in orbit, so bypassing the
limits of launch vehicles. The fog that is folding the future of
ISS and the general decline of funds for Space based Astronomical
Research, have suggested a less ambitious solution. The present
design is based on a telescope of completely new technology,
drastically reducing the weight to area ratio, with an effective
area of the order of 5 m$^{2}$ and a focal length of 35 m,
harbored on one satellite. A second satellite will be launched
with a complex focal plane instrumentation locked to the first in
a formation flight. The ensemble should be put in a L2 orbit.

The focal plane would be based on two major instruments, a
micro-calorimeter array based on TES technology and a Wide Field
Imager based on APS technology. These two instruments will be the
drivers for the study of remote Universe, explore the formation of
large scale structures, the role of black holes and their
interaction with galaxies. But they will provide as well the
capability to study deep Physics of more bright objects. To this
purpose the XEUS team is foreseeing the possibility to include
more instrumentations, mainly devoted to a more restrict
population of objects, but capable to address questions of extreme
interest. These auxiliary instrumentations to be implemented if
compatible with available resources and on the basis of a
trade-off of the science, include a photoelectric polarimeter
named XPOL.


\section{X-Ray Polarimetry}
\label{sect:title}

\subsection{Generalities}
In Optical Astronomy polarimetry is somehow a niche discipline. In
X-ray Astronomy it is expected to be a major topic for various
reasons:
\begin{itemize}
\item The non thermal emission processes, such as synchrotron,
cyclotron and non thermal bremsstrahlung play a significant role.
They result in highly polarized radiation. X-ray polarimetry can
enlighten the nature of the emission.
\item The transfer of radiation from the emitting region to the
observer undergo various processes that can result into a relevant
polarization.
\begin{enumerate}
\item In compact objects the radiation is scattered  by regions
whose geometry is far from spherical symmetry, including accretion
disks and accretion columns. The selection of scattering angle in
the direction of the observer result in a polarization, according
to the Compton/Thomson cross sections. The scattering can modify
the spectrum if the region is at high temperature or simply remove
a part of photons which are photoabsorbed. This last process of
scattering is usually named \textit{reflection} and is marked by
the presence of Fe K fluorescence photons (obviously unpolarized).
\item In the presence of very high magnetic fields the radiation
can be polarized by the combined effect of differential cross
sections for the scattering of normal and anomalous modes in the
plasma and vacuum birefringence, as predicted by Quantum
Electrodynamics.
\item In extended objects with thermal collisional equilibrium,
such as clusters of galaxies, if the plasma is thin for the
continuum and thick for resonance lines, the angle of last
scattering in the direction of the observer will be selective and
different in each part of the cluster resulting in a angular
dependent polarization of the line photons.
\item If some theories of Quantum Gravity based on the violation of
the Lorentz invariance are true the vacuum is birefringent and the
plane of polarization of photons should rotate of an amount that
increases with energy and with distance. The detection of this
effect would be of the highest interest. X-ray polarimetry can
perform this test at highest energies and on long distance base by
measuring the (expected) polarization of BL Lac objects and GRB
afterglows on the Gpc scale.
\end{enumerate}
More in general a vast literature, distributed on almost 40 years,
has made many predictions of the expected polarization from most
of classes of X-ray sources. Yet the only positive detection has
been, until now, that of Crab Nebula, first with a rocket then
with OSO-8 satellite. This is one of the brightest sources in the
x-ray sky, while many of the predictions refer to sources much
fainter.
\end{itemize}

\subsection{Technicalities}
The statistics for a counting rate polarimeter applies also to the
Micro Pattern Gas Detector at the focus of an X-ray optics. The
relevant statistical quantity is the Minimum Detectable
Polarization (MDP) which is the level of rejection of the
hypothesis of absence of polarization with a confidence level of
99 \% :

\begin{center}
              MDP = $\frac{4.29}{\mu S}$ ($\frac{S+B}{T}$)$^{0.5}$
\end{center}

where S is the source counting rate, B is the background, T the
observing time and $\mu$ is the modulation factor which is defined
as
\begin{center}
   $\mu$ = $\frac{C_{max}-C_{min}}{C_{max}+C_{min}}$
\end{center}

for a 100 \% polarized X-ray source, where $C$ is the counting
rate of the photoelectrons emitted in an a given angle-bin.

\section{SCIENCE WITH XEUS}
XEUS (X-Ray Evolving Universe Spectroscopy) it  will be proposed
as one of the major Missions to be included in the ESA Cosmic
Vision 2015-2025. If approved it could be the only large X-ray
telescope for many decades. Unavoidably, in this case, it will be
used as a multi-purpose observatory, to cover any observations, in
the X-ray band, of a certain level of interest. But the design of
the mission is based on choices which find their rationale in a
selected cluster of hot topics of Astrophysics and Fundamental
Physics that can be addressed with data from an X-ray observation.
The Major Goals of XEUS Mission are:

\begin{itemize}
\item Evolution of Large Scale Structure and Nucleosynthesis.
\item Formation, dynamical and chemical evolution of groups
and clusters.
\item Baryonic composition of the Intergalactic
Medium.
\item Enrichment dynamics: inflows, outflows and mergers.
\item Coeval Growth of Galaxies and Supermassive Black Holes.
\item Birth and Growth of Supermassive Black Holes.
\item Supermassive Black Hole induced galaxy evolution.
\item Matter under Extreme Conditions.
\item Gravity in the strong field limit.
\item Equations of state.
\item Acceleration phenomena.
\end{itemize}

These topics will be attacked with the primary instruments (NFI
and WFI). The addition of a Polarimeter can significatively
improve the science in some of these topics and add a few more.
Therefore it is now foreseen as an auxiliary instrument, to be
included compatibly with the tough technical constraints of such
an ambitious mission.


\section{Polarimetry for XEUS: capabilities and criticalities}
With XEUS optics X-ray polarimetry can be performed on celestial
sources with fluxes down to fraction of mCrab up to Crab level.
Very low MDP can in principle be reached with sources of very
large fluxes. The FOV of XPOL, included a fiducial region
determined by the extension of the tracks, is large enough to
include the principal X-ray features of the Crab Nebula and to map
most of the plerions with few pointings. Being capable to sustain
large fluxes, good control of systematic effects permits to
exploit at best the potentiality of the polarimeter. Photoelectric
polarimetry and the modern concept of a subdivided detector
permits to overcome most of the potential systematic effects
connected with the use of the classical techniques in space such
as non-uniformities in the Berillium window or in the analyzer. In
this case, the analyzer (and the detector) is the gas itself which
is intrinsically isotropic. The use of CMOS ASIC chip further
permits to maintain the control of systematic below 1 \% and at
level not yet reached by laboratory measurement. A measurement
which control the level of systematic effects  can be accomplished
by exploiting the internal calibration facility of the chip and
with the possible use of an unpolarized low-activity Fe$^{55}$
removable X-ray source. Background non-uniformities could be less
important in L2 orbit with respect to Low Earth orbit. However
they can be monitored exploiting the large field of view of XPOL,
during the observation, selecting different regions of the
detector outside the fiducial region containing the source in
observation.

\subsection {Data flow}
The data managing for XPOL will take into account of the source
rate under study. The requirements on the storage memory on board
and on the time required to download the data will therefore
remain limited. XPOL gather information from the GEM (Energy and
time) and from the ASIC chip (coordinates of the frame vertex and
energy and coordinates of the pixel in the frame). With the high
throughput of XEUS we will compress data on-board
(zero-suppression) by excluding the pixels with zero charge in
further storage/download. Ad example for a crab-like spectrum the
average rate of pixels before the zero-suppression is 630
pixel/event/s, after the zero-suppression the average rate of
pixel/events/s becomes 24.5. For sources with fluxes above 10
mCrab we implement an on-board analysis which reconstruct the
absorption point and photoelectron emission angle both based on
well known simple algorithms. The event will be tagged with those
information together with a quality parameter factor.  For the
Crab the expected bit rate is in this case 3.3 Mbit/s.

\section{The MPGC}
The Micro Pattern Gas
Detector\cite{costa01}$^,$\cite{Bellazzini06}$^,$\cite{Baldini06}
was first developed in 2001, with technology based on multilayer
printed circuit board (PCB) with vias which brought the signals
from the readout plane, patterned with hexagonal pixels 260 $\mu$m
pitch, to the readout electronics. A Gas Electron Multiplier
performs the multiplication of the charge produced by  the
photoelectrons tracks created in the drift/absorption region. The
PCB technique limited the number of channels (1000-2000) and the
pitch of the pixels (100-200 $\mu$m). Also long path to bring the
signals were capacitatively coupled one to each other. The jump in
the number of channels (hence in the coverage area and the FOV)
and the decreasing in the pixel pitch (hence in the response to
lower energy and the eventual use at high pressure) was it
possible by designing and building an ASIC CMOS multilayer chip.
The top metal layer is patterned in an honeycomb hexagonal pattern
which is the readout plane of the charge produced by the
photoelectrons. Each pixel is connected to a full electronics
chain (pre-amplifier, shaping amplifier, sample and hold,
multiplexer) built immediately below it, exploiting the remaining
five layers of the VLSI CMOS technology. Three generations of
chips have been developed so far with increasing area and
throughput. The first VLSI CMOS chip (chip I) having 4 mm
diameter, 2101 pixel chip and 80 $\mu$ pitch, was provided with
external trigger and downloading of all the frame pixels with a
rate capability of about 1 kHz. The second generation of chips,
squared 1.1 cm x 1.1 cm (chip II), was patterned with 22000 pixels
organized in 8 independent sub-frames of 2700 pixels with external
or internal trigger. The frames can be readout in parallel
providing a frame rate of 1-2 kHz. The third, and current
generation of chip (chip III) is provided with unique
characteristics to be used with the best performances for X-ray
polarimetry.

\subsection{The 105600 pixel chip and the Monte Carlo simulation.}

The chip, which performances are described in detail in
\citenum{Bellazzini06a} has 105600 hexagonal pixels arranged at 50
$\mu$m pitch in a honeycomb matrix, corresponding to an active
area of 15$\times$15 mm$^{2}$. Each pixel (as the precedent
generations) is connected to a charge-sensitive amplifier followed
by a shaping circuit and a sample and hold-multiplexer circuit.
The chip is organized in 16 identical clusters of 6600 pixels or
alternatively in 8 clusters of 13200 pixels each one with an
independent differential analog output buffer. In this last
generation however each cluster has a customizable internal
self-triggering capability with independently adjustable
thresholds. Every 4 pixels contribute to a local trigger with a
dedicated amplifier. The chip is capable to select for each event
the rectangular region of interest (ROI) containing the track.
This reduces the time of readout and the rate can be as high as
20~kHz. Also the noise is very low, 50 e$^{-}$/pixel, therefore
the single electron readout is possible also with a small GEM
gain.

Monte Carlo simulations have been developed on the base of well
established functions that describe the interactions of electrons
in ambient media and which are specialized for low energy
applications and for gas mixtures. The simulations include the
generation of `s' and `p' photoelectrons with the correct weight
and initial angular distribution derived from the interaction of a
polarized and unpolarized X-ray photon with the atoms of the gas.
The performance for different configurations at different gas
mixtures and pressure and thickness as a function of energy can be
studied therefore in detail. The configuration adopted for XEUS is
conservative. Micro Pattern Gas Detector with thickness of 1 cm,
which provides negligible blurring of the image at the focus of
the converging beam from the 35 m focal length mirror optics, and
pressure of 1 atmosphere are currently in use in our laboratory.

\section{The Detector for XPOL}
\begin{figure}
\begin{center}
\begin{tabular}{c}
\subfigure{\includegraphics[angle=0,width=11.0cm]{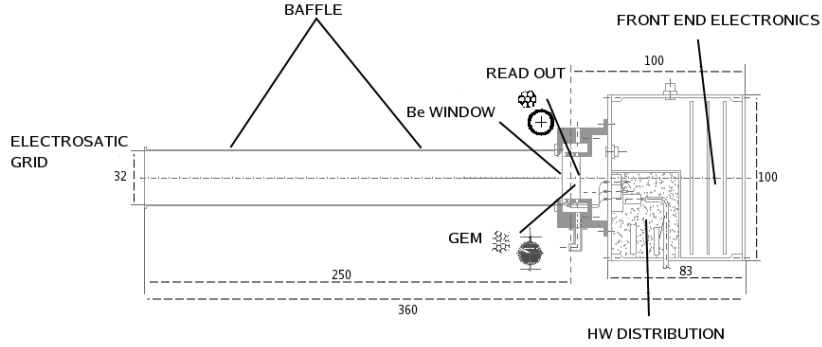}}
\end{tabular}
\end{center}
\caption[example]{Design of the X-ray Polarimeter which includes
the baffle to limit the field of view of XPOL by the mirror optics
aperture and prevent diffuse X-ray background coming from the
space which separates the two spacecraft modules to interact with
the detector of XPOL (quotes are in mm).} \label{Xpol}
\end{figure}

Hereafter we show the main configuration properties for the
detector XPOL shown in fig. \ref{Xpol} which basically consists of
a well established technology gas cell with a charge multiplier
coupled with a modern ASIC VLSI chip.

Since it provides simultaneously imaging, spectra and
polarization, Micro Pattern Gas Detector could be the long sought
laking tool for X-ray astronomy. Nowadays the use of CMOS VLSI
chip permits to reach field of view of 1.5' x 1.5' at the focus of
35m XEUS optics therefore bringing XPOL already at the same level
of throughout of the other instruments.

\begin{table}[h]
\caption{Main characteristic of XPOL. In parenthesis are shown the
resolution of the optics which drive the overall position
resolution. } \label{tab:charact}
\begin{center}
\begin{tabular}{|l|l|} 
\hline
\rule[-1ex]{0pt}{3.5ex}  Position Resolution (FWHM, Linear)& 150 $\mu$m (900 $\mu$m)  \\
\hline
\rule[-1ex]{0pt}{3.5ex}  Position Resolution (FWHM, Angular) & 0.88'' (5'')   \\
\hline
\rule[-1ex]{0pt}{3.5ex}  Field of View (Linear)& 1.5 cm $\times$ 1.5 cm  \\
\hline
\rule[-1ex]{0pt}{3.5ex}  Field of View (Angular) & 1.5' $\times$ 1.5'  \\
\hline
\rule[-1ex]{0pt}{3.5ex}  Maximum Asynchr. Readout & 10-20 kHz  \\
\hline
\rule[-1ex]{0pt}{3.5ex}  FWHM @ 6 keV & 15\%  \\
\hline
\rule[-1ex]{0pt}{3.5ex}  Timing accuracy & 2 $\mu$s \\
\hline
\end{tabular}
\end{center}
\end{table}

\subsection{Gas Cell}
The gas cell is designed to include:
\begin{itemize}
\item A beryllium window 50$\mu$m thick.
\item A set of field forming rings.
\item A frame supporting the GEM.
\item The VLSI chip described in section 5.1.
\item Feed-through to carry bias voltages to the various internal
components.
\item A ceramic case to contain the gas.
\item A ceramic base to mount the chip and route outside the
signals and controls for the chip.

\end{itemize}

\subsection{Gas Mixture}
We do not discuss in detail the choice of the optimal filling
mixture for the detector that will be object of further
investigations. Off the recipe book of mixtures simulated so far
we select the mixture He (40$\%$) DME (60$\%$) that seems well
tuned to the present baseline design of the XEUS telescope and has
the significant advantage of having been already tested with
excellent results, coherent with what predicted by the Monte Carlo
simulations. This makes the predictions we show in the next
paragraphs as realistic as compatible with such an ambitious
experiment. We stress, anyway, that this is far from any fix
choice and a large room of improvement is there.

\section{XPOL performance} \label{sect:astro}

\subsection{Efficiency and Modulation factor}

\begin{figure}
\begin{center}
\begin{tabular}{c}
\includegraphics[angle=0,width=8.2cm,angle=90]{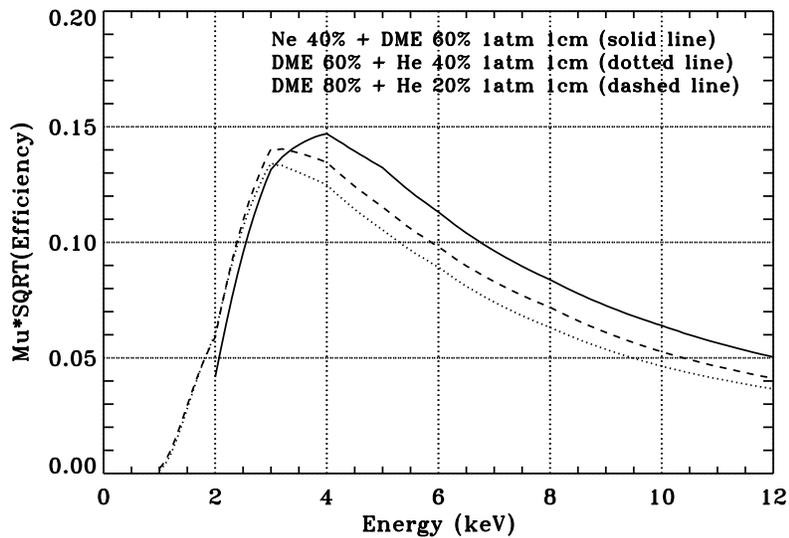}
\end{tabular}
\end{center}
\caption[example]{Quality factor derived from simulations for
three tested mixture to be possibly used with XPOL on-board XEUS.
He-DME mixtures provides the better polarimetric response to the
XEUS optics while a Ne-DME mixture permits a better performance at
the higher energy end.} \label{fig: Gas perf}
\end{figure}

Being a focal plane imaging instrument filled with low-Z gas (DME,
He or Ne) XPOL has an intrinsic very low background
\cite{Bunner78}. Working down to 1.5-2 keV, in a large energy
band, it overcomes the limits of the traditional approach of
scattering polarimeter, which are insensitive below 5 keV and are
background limited, and of Bragg crystal polarimeter, which work
in a narrow band around the order of diffraction. The sensitivity
of XPOL, therefore, is mostly dominated by the source flux down to
very low values, therefore the sensitivity at a certain energy is
proportional to the product $\epsilon ^{0.5} \times\mu$ where
$\epsilon$ is the efficiency of the gas mixture and $\mu$ is the
modulation factor above defined. This is the quality factor of the
polarimeter which drives the choice of the main physical parameter
such as the drift region, the gas mixture and pressure. The choice
of a gas mixture defines the energy band in which the X-ray
polarimeter operates. We present here the compared quality factors
for 1-cm 1-Atm for low atomic number mixtures. The result shows,
basically, that the mixtures based on He-DME, which is currently
proposed for the polarimeter on XEUS, better exploits the high
effective area of the mirror at low energy. Nevertheless the
polarimetric response at higher energies aimed to study the
physics of the scattering privilege Neon-based mixtures. For now
we consider the tested mixture (He 40 $\%$-DME 60 $\%$) which
assures the better overall performance in terms of Minimum
Detectable Polarization. From the simulations results that already
an improvement at higher energies without loosing sensitivity at
low energy derives from an increasing of the percentage of DME
(see fig. \ref{fig: Gas perf}).

\subsection{Interface Electronics}
\begin{figure}
   \begin{center}
   \begin{tabular}{c}
   \includegraphics[height=7cm,angle=0]{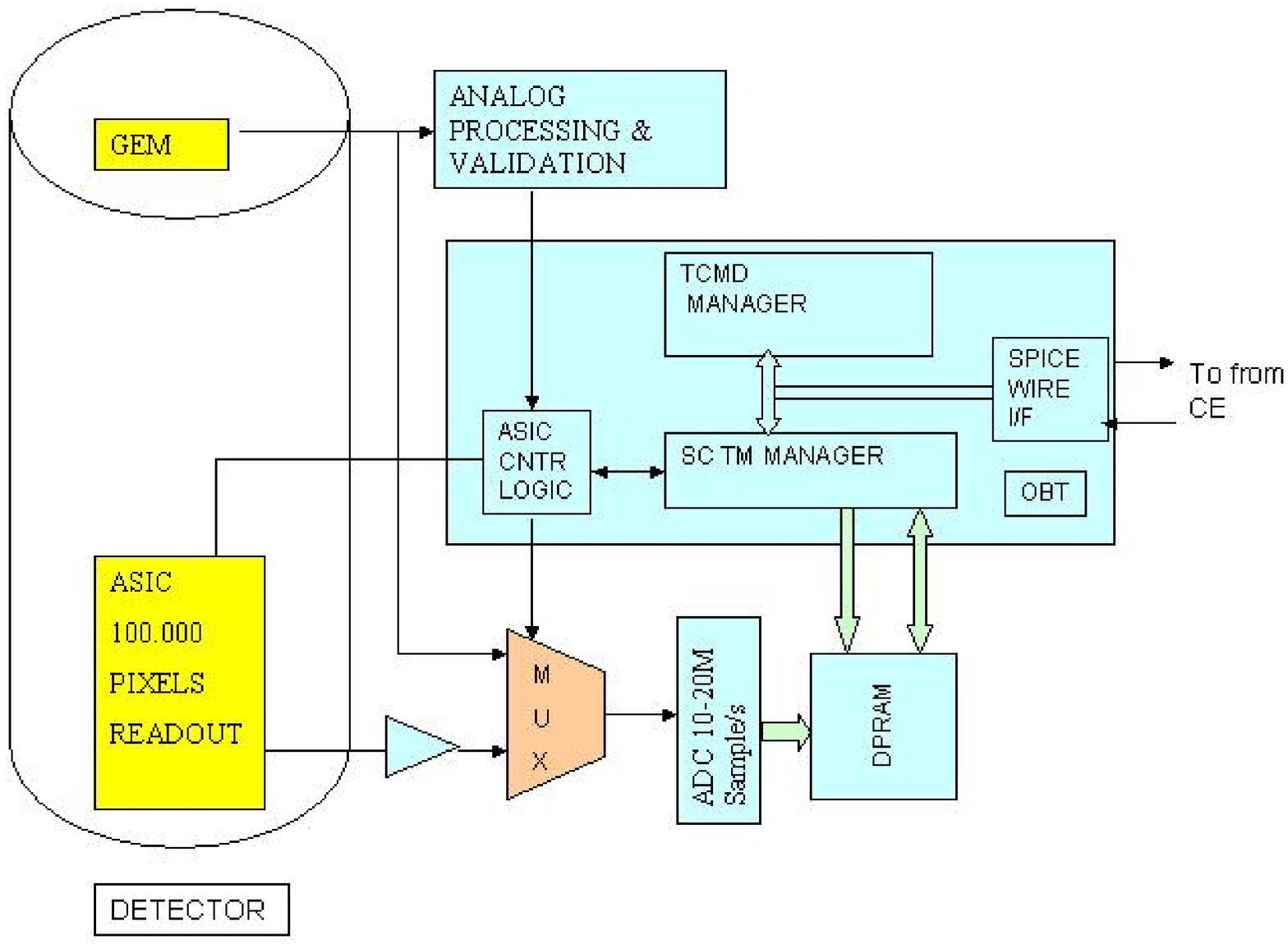}
   \end{tabular}
   \end{center}
   \caption[example]
{\label{fig:example} The Interface electronics and the Data
Handling will be capable to manage the high troughput data from
XEUS } \end{figure}

The interface electronics (see fig. \ref{fig:example}) will be
capable to perform the following functions:
\begin{itemize}
\item Configure the ASIC CMOS.
\item Process the data from the Gas Electron Multiplier.
\item AD convert  the pixel energy content.
\item Store the pixel in the memory and perform the
zero-suppression.
\item Organize the event data structure.
\end{itemize}
Most of the characteristics typical of a front-end electronics are
performed by the ASIC chip. Therefore the architecture will not
require exotic solutions but will be mainly dedicated to configure
and readout the data from the chip. The power consumption and the
dimension therefore will be contained (see tab.
\ref{tab:resource}).
\begin{table}[h]
\caption{ Weight volumes and power consumption for XPOL. FEE is
the Front-End electronics, CE is the control Electronics which
manage the telecommand to configure the chip, provide the mass
memory to store the data and provide the power for the FEE and the
Micro Pattern Gas Detector.} \label{tab:resource}
\begin{center}
\begin{tabular}{|l|l||l|l|} 
\hline
\rule[-1ex]{0pt}{3.5ex}  Unit & Mass (Kg)& Power (W) & Dimension (cm) \\
\hline
\rule[-1ex]{0pt}{3.5ex}   Detector + Baffle & 2.0 & 3.5 & 28 $\times$ 6 \O \\
\hline
\rule[-1ex]{0pt}{3.5ex}      FEE & 1.5 & 3.5 & 10 $\times$ 8 $\times$ 10 \\
\hline
\rule[-1ex]{0pt}{3.5ex}   CE & 7.5 & 10 & 27 $\times$ 15 $\times$ 20  \\
\hline
\rule[-1ex]{0pt}{3.5ex}   Total & 11.0  & 17.0 & 10 $\times$ 8 $\times$ 10 \\
\hline
\end{tabular}
\end{center}
\end{table}

The AD conversion will be performed at the maximum rate compatible
with the ROI frame readout which at the moment is 10-20 Mhz. Pixel
with no signal will be suppressed. Timing relies on the fact that
the signal from the GEM is very fast and can be shaped by less
than 1 $\mu$s without sensitive loss of signal. However most of
the uncertainty is determined by the drift time in the gas which
can be of the order of 1 $\mu$s. The overall accuracy will
therefore be of $1.5-2 \mu$s.

\subsection{Sensitivity}
The sensitivity of XEUS takes advantage of the large area and the
good PSF. The contribution of the diffused X-ray background result
therefore negligible down to the flux level for which polarization
measurements are feasible in an observing time compatible with the
planning of XEUS (10 $\mu$Crab). The internal (residual)
background is expected to be very small. Particle background will
be discriminated by the high granularity of the detector and by
the fiducial sides. Compton electrons, releasing energy in the
operative range of XPOL, are few because of the low atomic number
of the gas exploited. In our calculation we conservatively do not
apply any further background-rejection, even so the particle
background is negligible.

\begin{figure}
\begin{center}
\begin{tabular}{c}
\subfigure {\includegraphics[angle=0,width=10.0cm]{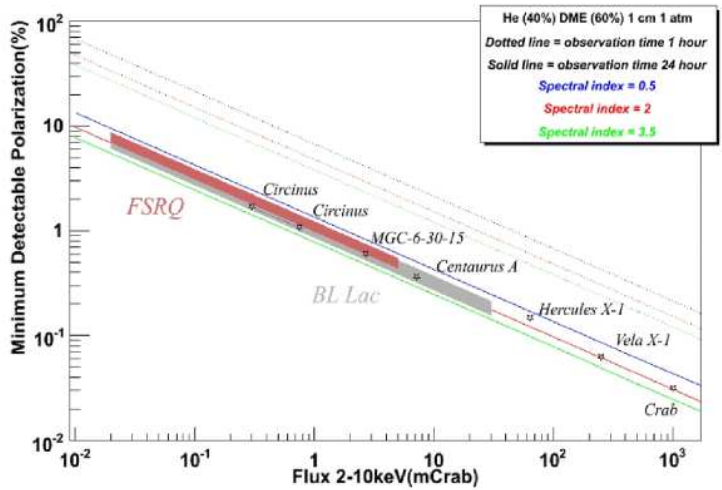}}
\end{tabular}
\end{center}
\caption[example]{Minumum Detectable Polarization for XPOL. Very
high sensitivity make XPOL to become the high throughput tool for
X-ray Astronomy} \label{fig:MDP_xeus}
\end{figure}


\begin{figure}
\begin{center}
\begin{tabular}{c}
\subfigure[\label{fig:gal_center}]{\includegraphics[angle=0,width=8.2cm]{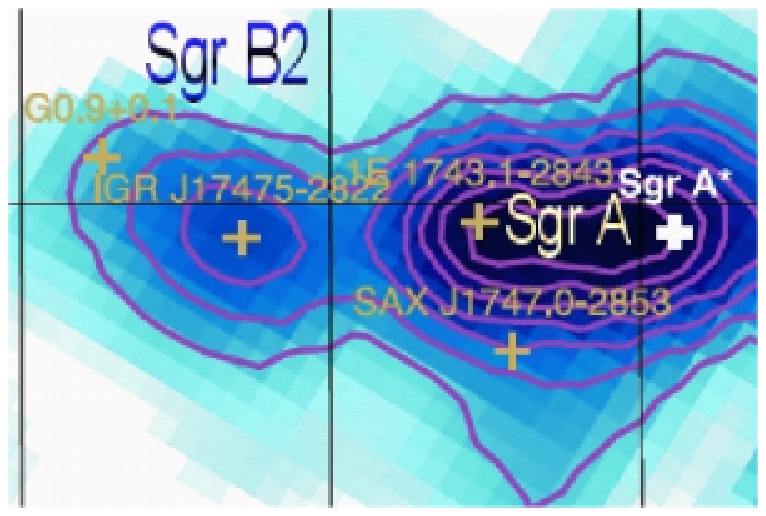}}
\subfigure[\label{fig:pol_scatt}]{\includegraphics[angle=0,width=8.2cm]{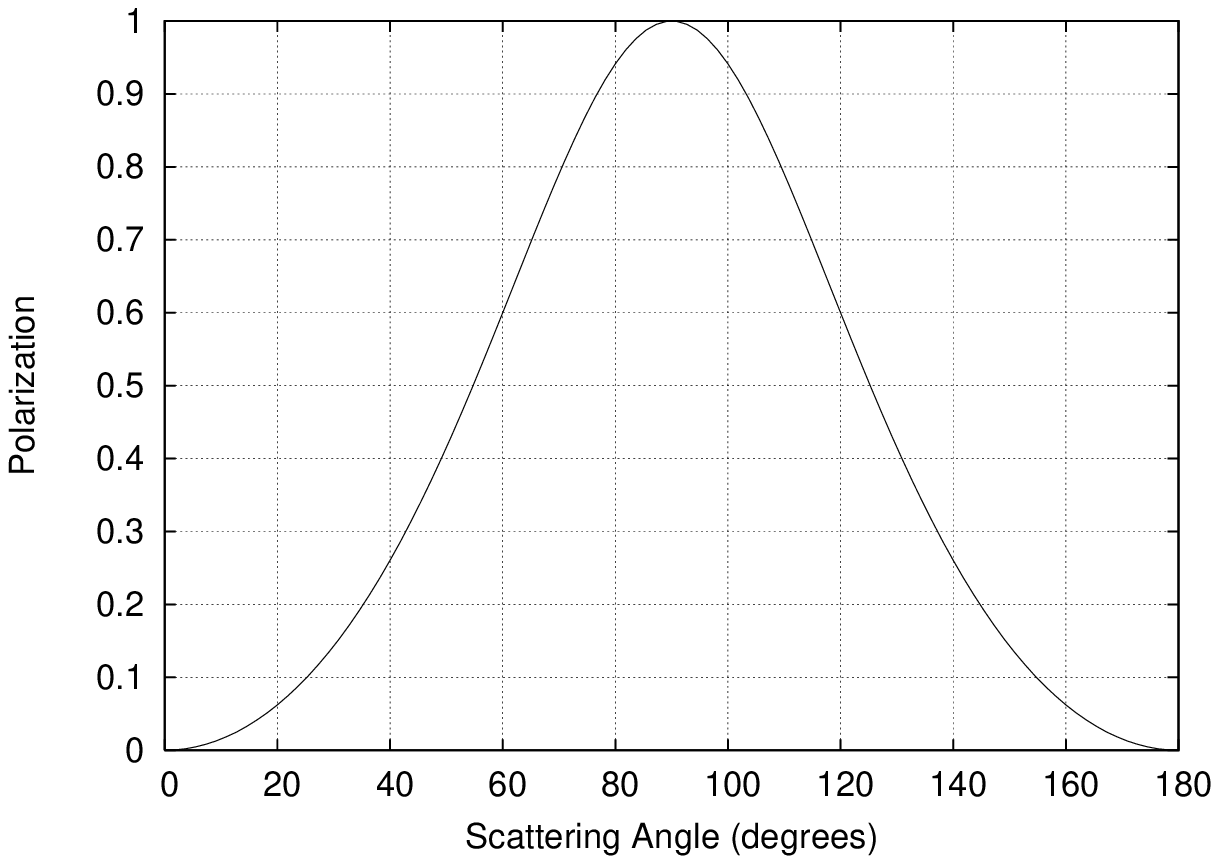}}
\label{fig:accelerator}
\end{tabular}
\end{center}

\caption[example]{Panel (a) shows part of the Galactic Center
region from the INTEGRAL observation \cite{Revniv04}. The giant
molecolar cloud Sgr B2 probably is reprocessing the X-ray
radiation coming from Sgr-A* since 300-400 years. From the degree
of polarization of scattered radiation (Panel (b)), one can
estimate the angle of scattering and, thus, the correct distance
of Sgr B2 from Sgr A*.}

\label{fig:NeDME8020}
\end{figure}

\section {Some goals for XPOL}

\subsection{Was The Milky way an AGN in the recent past?}
The region around the Galactic Center is complex. There are
molecular clouds for which is only known the projected distance
respect to us. One of these molecolar clouds, Sgr B2, has been
studied \cite{Koyama96} in detail by ASCA  and more recently by
INTEGRAL \cite{Revniv04}. The spectrum shows an iron line and a
reflection component. Also it looks brighter on the direction of
Sgr A* which is supposed to harbour a massive black hole.  The
prevalent (and old) hypothesys\cite{Sun93} is that Sgr B2 is
reprocessing the radiation pervening from the central black hole.
The latter was flaring 300-400 years ago which is the travelling
time to Sgr B2. The measurement \cite{Chura02} of the degree and
of the polarization angle will permit to confirm this hypothesis
and to measure the real distance to the central BH providing an
estimation of the luminosity of Galactic Center when it was a
faint AGN in the past. In one day of observation of XPOL the MDP
expected is 6.8 \%.

\subsection{Matter in extreme Magnetic fields: testing QED}
The presence of magnetic fields of the order of $10^{13}$,
$10^{14}$ gauss in neutron stars such as Anomalous X-ray Pulsars
or Soft Gamma Ray Repeater (SGR) are mainly derived on the base of
energetic considerations. QED predicts that, at this large
magnetic field value,  the vacuum becomes birefringent and that
the polarization degree from magnetized neutron star can assume a
large value.\cite{Laho03}$^,$\cite{hesha02} Due to the extreme
magnetic field also two absorption features should be present, one
at the so called vacuum resonance frequency, and one at the
proton-cyclotron resonance\cite{Nimbul06}. These features should
have different phenomenology with respect to polarization. This
would provide a direct evidence of the presence of extreme
magnetic fields and provide a check of the magnetar model. In an
active phase or in the first few days following a major emission
episode the situation may become even more interesting, because of
the likely presence of a transient ionized atmosphere. The
presence of a feature at 5~keV in the spectrum of
SGR1806-20\cite{Ibra02}  has been explained as due to a proton
cyclotron line \cite{zane01}. The spectroscopic capability of the
Micro Pattern Polarimeter allows to perform energy resolved
polarimetry and test the QED predictions. The lack of QED effect
on polarization would imply the presence of a red-shifted iron
line which has deep implication in the determination of the
Equation of State of the neutron stars.

\subsection{Matter under extreme Gravitational Fields: testing GR}
Radiation emitted from the innermost region of an accretion disk
can be polarized by means of scattering. At a distant observer,
neglecting General Relativity (GR) effect, the polarization vector
is parallel or perpendicular to the major axis of the projected
disk on the sky \cite{LigShap76}. In presence of a Black Hole in a
soft-state the disk is optically thick and geometrically thin
extending down in the vicinity of the BH, when GR is taken into
account, a continuous rotation of the polarization vector with
energy is expected \cite{Connorsstark77}$^,$\cite{CoStaPi80}. In
fact photons more energetic are emitted closer to the BH and the
rotation due to effect of GR is therefore larger.

\begin{figure}
\begin{center}
\begin{tabular}{c}
\subfigure[\label{fig:Cyg_X1mod}]{\includegraphics[angle=0,height=7.0cm]{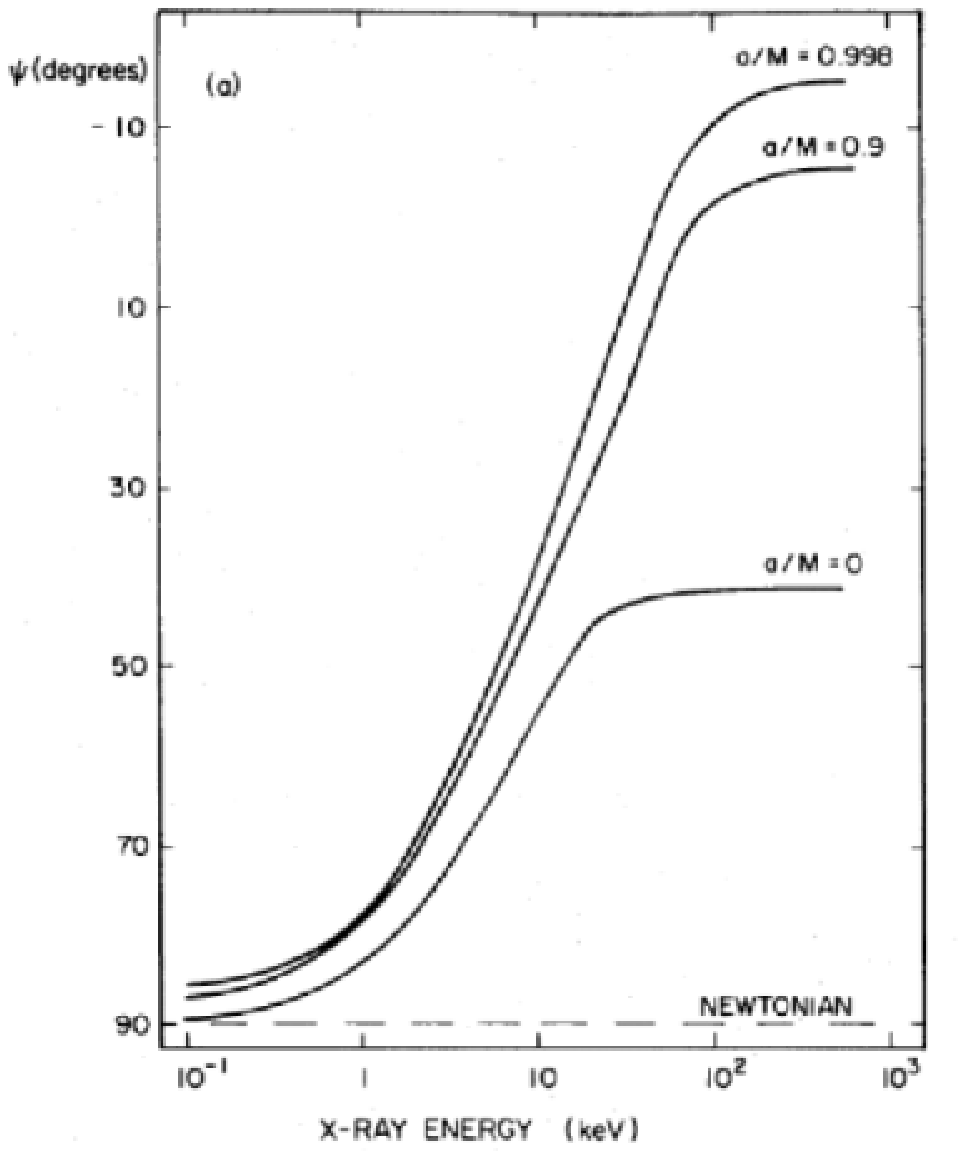}}
\subfigure[\label{fig:CygX1sim}]{\includegraphics[angle=0,height=7.0cm]{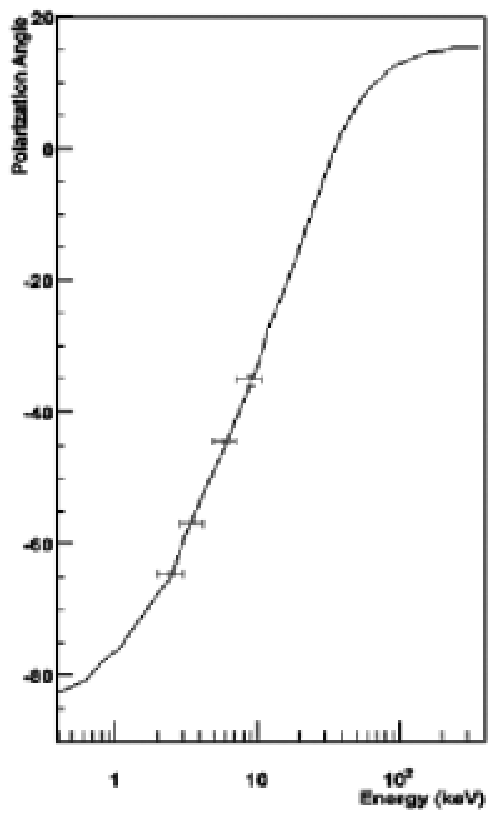}}
\end{tabular}
\end{center}
\caption[example]{(a) Expected variation of the polarization angle with energy of
the radiation coming from the accretion disk of a
BH\cite{Connorsstark77}$^,$\cite{CoStaPi80} (b) result of a simulated observation
of 12 hours of Cyg X-1 where the plane of polarized (5\%) radiation is rotated
according to the prescription of (\citenum{Connorsstark77}). We can observe the
rotation with a very high statistical significance (see text).} \label{fig:BH}
\end{figure}

We simulated an observation for Cyg X-1 in soft state of 12 hours
with XEUS and in the input data (we supposed a polarization of 5 
we rotated the plane of polarization according to the prescription
of \cite{Connorsstark77}. The results are shown in figure
\ref{fig:BH}. Clearly the rotation is detected with a far high
statistical significance. XPOL will be able therefore to detect
such a variation either with a much less integration time (the
error scale as $\sqrt{T}$) and with a lower intrinsic polarization
of the source. More with good statistical significance would be
possible to detect the momentum (maximally rotating/non rotating)
of the BH.

\subsection{Cosmic Accelerators: the prototype}
\begin{figure}
\begin{center}
\begin{tabular}{c}
\subfigure[\label{fig:accel_gal}]{\includegraphics[angle=0,height=6.0cm]{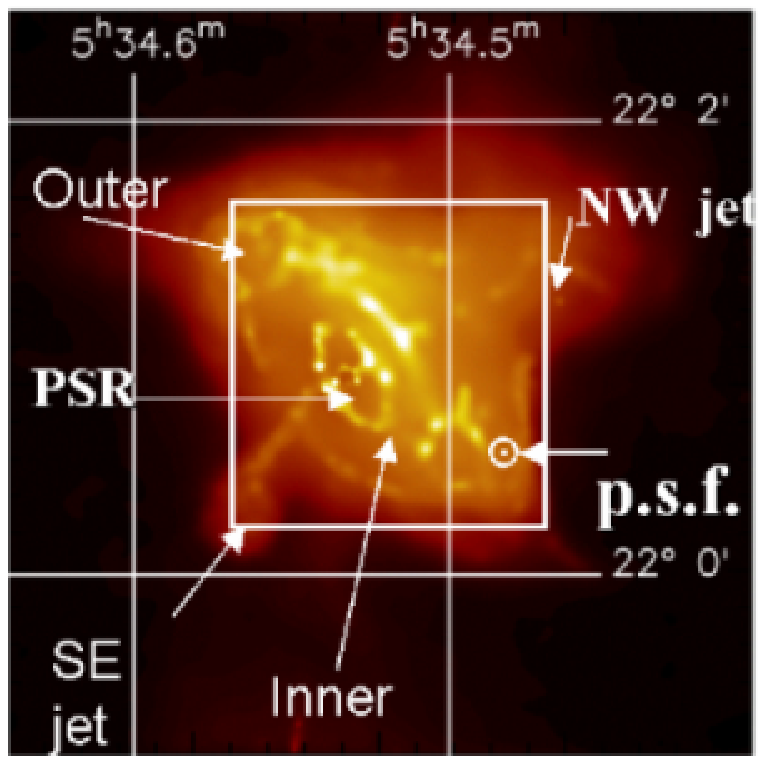}}
\subfigure[\label{fig:accelextragal}]{\includegraphics[angle=0,height=6.0cm]{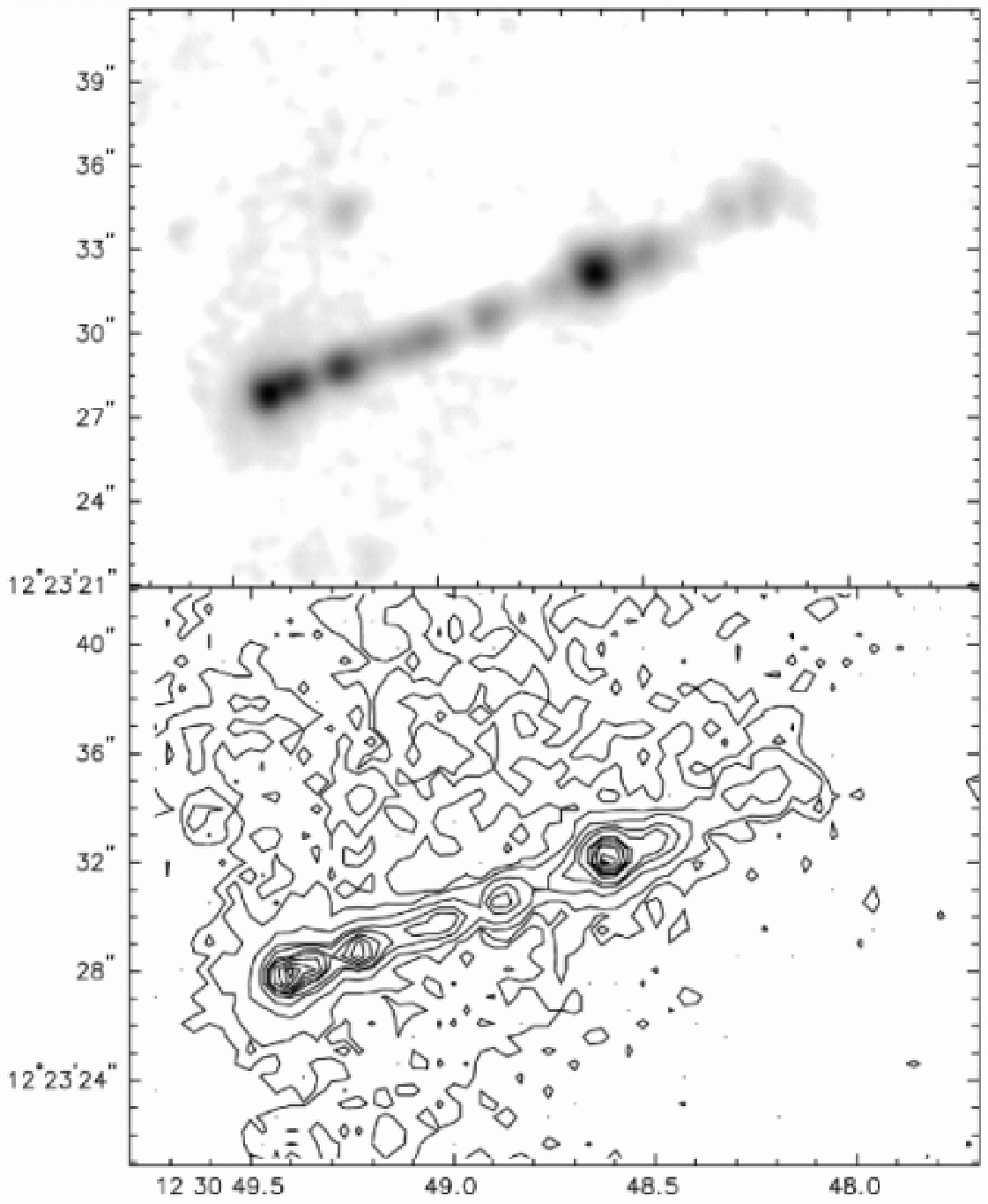}}
\end{tabular}
\end{center}
\caption[example]{Crab Nebula Observed by Chandra ({\bf a}) with
overimposed the field of view (FOV) of XPOL, the PSF of the XEUS
mirrors and, inside, the PSF of the polarimeter, {\bf b} grey
scale of the M87 jets (top) observed by Chandra and the brightness
contour (bottom) with the core on the leftest brightest peak and
the knots on the right.} \label{fig:accelerators}
\end{figure}

Even if considered to be the calibration source of X-ray sky
observatories, Crab Nebula continues to reserve many surprises
(see fig. \ref{fig:accel_gal}) . When observed with the fine
position resolution of Chandra \cite{Weisskp00}, many
substructures are evident:  an inner torus, an outer torus and two
jets . The standard picture of the Crab Nebula was derived long
ago by \citenum{Kencor84}. Kennel and Coroniti assumed that a
magnetohydrodynamic wind from the pulsar terminates in a shock and
the post-shock shines in the synchrotron nebula. In their model
the frozen magnetic field is supposed to be toroidal. However if
this is the case the brightness around the torus will not be
uniform in two opposite direction of the torus. The hypothesis
that the magnetic field be, instead, turbulent resolves this
discrepancy and can be probed by polarization angle measurements
resolved in space which are possible with XPOL thanks to its
spectral/imaging capability. By mapping the polarization angle on
the Crab image will be possible to gather information on the
spatial structure of the magnetic field and on the physics of the
interaction of the wind with the nebula.

\subsection{Cosmic Accelerators: the jets}
Jets in Blazars, radio galaxies and in Micro Quasars are the sites
of acceleration of particle up to very high energies which can
reach also TeV. In most of the cases the particles (leptons) can
be responsible of the observed X-ray radiation. In Blazars low
energy photons can be up-scattered by the same emitting electrons
(Synchrotron Self-Compton) to X-rays or, conversely, the external
UV photons from the disk can be Comptonized to X-rays. The
polarization degree and angle with respect to other wavelengths is
different for this two cases. \cite{Celmatt}$^,$\cite{Pout94}
X-ray polarimetry therefore permits to disentangle between the two
models establishing the origin of the emission, the role of the
disk and the nature of the jet. In Radio Galaxies, like the FR I
M87, and in the case of Micro Quasar XTE J1550-564 the Jet
interacts with the ambient medium and forms bright knots at a
distance from the central core. The X-ray coming from those knots
can be resolved (see fig. \ref{fig:accelextragal}) and analyzed by
XPOL to determine if the degree of polarization is compatible with
a synchrotron origin. Conversely, if X-rays are generated by
comptonization of external photons for example those deriving from
Cosmic Microwave Background Radiation, the degree of polarization
may be very small for the hysotropy of the source photons.



\section{CONCLUSIONS} \label{sect:conclusions}
The advent of the Micro Pattern Gas Detector, a finely subdivided
pixel gas chamber detector readout by large area and small pixel
size CMOS ASIC  allows to image the track of the photoelectron and
derives  the polarization of the X-ray photon together with energy
position and time. XPOL, as auxiliary instrument at the focus of
the large mirror area of XEUS telescope, will be decisive in many
of the scientific goal of the mission being therefore the high
throughput no more laking tool of X-ray Astronomy.
\acknowledgments     
This research could benefit of support from INFN, INAF, ASI. We
thank Alcatel Alenia Space LABEN division for support in onboard
data handling.

\bibliography{xeusreport}   
\bibliographystyle{spiebib}   
\end{document}